\documentclass[%
 reprint,
superscriptaddress,
 amsmath,amssymb,
 aps,
floatfix,
]{revtex4-2}
\usepackage{layouts}
\usepackage{graphicx}% Include figure files
\usepackage{dcolumn}% Align table columns on decimal point
\usepackage{bm}% bold math
\allowdisplaybreaks

\newcommand{\mc}{\mathcal}

\newcommand{\bra}[1]{\left\langle #1 \right|}
\newcommand{\ket}[1]{\left| #1 \right\rangle}

\begin{document}

\preprint{APS/123-QED}

\title{Ultrafast dephasing in solid state high harmonic generation: macroscopic origin revealed by real-space dynamics}

\author{Graham G. Brown}
\email{brown@mbi-berlin.de}
\affiliation{Max Born Institute, Max-Born-Stra\ss e 2A, 12489, Berlin, Germany}
\author{\'{A}lvaro Jim\'{e}nez-Gal\'{a}n}
\affiliation{Max Born Institute, Max-Born-Stra\ss e 2A, 12489, Berlin, Germany}
\affiliation{Instituto de Ciencia de Materiales de Madrid (ICMM), Consejo Superior de Investigaciones Cient\'{i}ficas (CSIC), Sor Juana In\'{e}s de la Cruz 3, 28049 Madrid, Spain}
\author{Rui E. F. Silva}
\affiliation{Max Born Institute, Max-Born-Stra\ss e 2A, 12489, Berlin, Germany}
\affiliation{Instituto de Ciencia de Materiales de Madrid (ICMM), Consejo Superior de Investigaciones Cient\'{i}ficas (CSIC), Sor Juana In\'{e}s de la Cruz 3, 28049 Madrid, Spain}
\author{Misha Ivanov}%
\affiliation{Max Born Institute, Max-Born-Stra\ss e 2A, 12489, Berlin, Germany}
\affiliation{Department of Physics, Humboldt University, Newtonstra\ss e 15, 12489 Berlin, Germany}
\affiliation{Blackett Laboratory, Imperial College London, London SW7 2AZ, United Kingdom}

\date{\today}

\begin{abstract}
Using a fully real-space perspective on high harmonic generation (HHG) in solids, we examine the relationship between microscopic response, macroscopic propagation of this response to the far field, and the extremely short dephasing times routinely used in the theoretical simulations of experimentally measured solid-state HHG spectra. We find that far field propagation naturally reduces the contribution to the observed HHG emission from electrons that do not return to the lattice site where they have been injected into the conduction band. We then show that extremely short dephasing times routinely used in microscopic simulations suppress many electron trajectories that contribute to the far-field spectra, leading to significant distortions of the true high harmonic response. We show that a real-space based dephasing mechanism, which preferentially suppresses trajectories which veer too far away from their original lattice site, yield HHG spectra that faithfully retain those trajectories that contribute to the far-field spectra while filtering out those which do not, already at the microscopic level. Our findings emphasize the similarities between atomic and solid-state HHG by highlighting the importance of the intensity-dependent phase of HHG emission and address the longstanding issue regarding the origin of extremely short dephasing times in solid-state HHG.
\end{abstract}

\maketitle

\section{Introduction}

For several decades, high harmonic generation (HHG) and attosecond spectroscopy have proven to be invaluable techniques for exploring ultrafast electron dynamics in atoms and molecules  \cite{doi:10.1126/science.1189401,doi:10.1126/science.1254061,CEDERBAUM1999205,Chini2014,NISOLI200917,RevModPhys.81.163}. Over the past decade, HHG spectroscopy has been expanded to investigate condensed matter systems and the breadth of topics available to solid-state HHG spectroscopy is far-reaching. These topics include studies of the band structure, both in the absence of external fields \cite{allOpticalBandStructure} and under laser excitation \cite{Uzan-Narovlansky2022}, density of states \cite{Uzan2020,PhysRevLett.118.087403}, topological edge states \cite{Baykusheva2021,PhysRevLett.120.177401}, electron-hole dynamics \cite{EH1,EH2}, multi-electron dynamics \cite{hhgManyBody,PhysRevA.105.053118}, Berry curvature \cite{HHGBC1,HHGBC2}, high-$T_c$ superconductivity \cite{https://doi.org/10.48550/arxiv.2201.09515}, and topological phase transitions \cite{HHGTPT,PhysRevX.13.031011}. 

Despite the conceptual differences between the formalisms used to describe atomic and condensed matter systems, atomic and solid-state HHG bear striking similarities. In particular, HHG in both atomic systems and condensed matter is described using a three-step model of recollision \cite{PhysRevA.49.2117,PhysRevLett.70.1599,PhysRevLett.71.1994,Kulander1992,vampa1}, for which the latter can be considered a generalization of the former to non-parabolic electron dispersion relations: (1) an electron-hole pair is created when an electron tunnels from a bound to an excited state in the presence of a strong field; (2) the electron-hole pair is accelerated by the strong field; (3) the electron-hole pair recombines and emits a high-energy photon. Like for atomic HHG, the three-step model of solid-state recollision has led to the development of semi-classical models which describe HHG emission using semiclassical trajectories of the electron-hole pair \cite{vampa1,imperfectRecollisions} and the spectral characteristics of HHG radiation is understood through these semiclassical models \cite{vampa1,vampa2,allOpticalBandStructure}. This real-space trajectory perspective is especially important for simulations of atomic HHG using long wavelength driving lasers at the microscopic level, where absorbing boundaries are used to suppress recollision trajectories whose high harmonic emission phase exhibits large intensity-dependence and is not observed experimentally due to macroscopic propagation effects \cite{Zhang2016,Rothhardt_2014}.

There are also important qualitative differences between atomic and solid-state HHG which we discuss with a real-space perspective. The periodic nature of condensed matter systems allows for radiative transitions between states separated by large distances within a crystal lattice \cite{WQC,imperfectRecollisions,wannierHHGLewenstein}. In atomic HHG, components of the continuum electron wave packet which never return to their parent ion lead to ionization, negligibly contribute to HHG spectra, and are typically removed from simulations through the use of an absorbing boundary. In contrast, the large cross-section for transitions between distantly separated states in periodic systems results in significant contributions to calculated solid-state HHG spectra from trajectories which never return to their parent lattice site \cite{brown2023realspace}.

These comparatively complex dynamics of solid-state HHG are obfuscated by the requirement for extremely short dephasing times $T_2 \sim 1$ fs, which are necessary to obtain agreement between experimentally measured and microscopically simulated HHG spectra. The physical significance of such ultrafast dephasing times has remained unclear \cite{RevModPhys.90.021002}, although important steps towards resolving this issue have been made in \cite{PhysRevA.97.011401,PhysRevLett.125.083901,Abadie:18}. No proposed microscopic dephasing mechanism, however, is both sufficiently strong and applicable to all materials undergoing recollision including atomically thin monolayers \cite{HHGBC2}. 

In a recent work \cite{brown2023realspace}, we developed a model of solid-state HHG formulated entirely in real-space using the Wannier basis \cite{wannierOG} and showed that the suppression of trajectories which never return to their parent lattice sites using a spatially-dependent dephasing mechanism results in clear HHG spectra which agree with far-field spectra calculated without ultrafast dephasing times. Here, we expand upon this work and use our real-space model of solid-state HHG to show that far-field propagation suppresses recollision trajectories which never return to their parent lattice site, that extremely short dephasing times artificially suppress recollision trajectories which return to their parent lattice site and contribute to far-field spectra, and that a spatial-dephasing mechanism which preferentially suppresses coherences between distantly separated states retains these trajectories which return and contribute to far-field spectra. We do this by calculating the dynamics of the electron-hole wave packet entirely in real-space, thereby allowing us to observe recollision trajectories contributing to HHG emission directly from numerical simulation.

Our results bridge a gap between atomic and solid-state HHG, wherein complex absorbing potentials are ubiquitously used in simulations of the former to mimic macroscopic experimental conditions. In particular, our results identify which real-space recollision trajectories survive far-field propagation in solid-state HHG and illustrate a means for suppressing HHG emission from trajectories which do not contribute to far-field spectra from microscopic simulation in a way completely analogous to the use of complex absorbing potentials in simulations of atomic HHG. 

\section{Describing Solid-State HHG in the Wannier Basis}
\label{section:NumericalModel}

We consider a two-band one-dimensional periodic system with lattice constant $a_0 = 4.2$ \AA (8 a.u.) with a Mathieu-type potential with strength 10 eV ($0.37$ a.u.) \cite{hhgFromBEiS}. We assume that the ground state has been found in the Bloch basis such that the Bloch basis wavefunction $\ket{\psi_{m, k}}$ in band $m$ with crystal momentum $k$ has energy $\epsilon_{m}^{k}$. We let $p_{m, m'}^{k}$ and $d_{m, m'}^{k}$ denote the momentum and dipole matrix elements between Bloch states in bands $m$ and $m'$ with crystal momentum $k$.

The Wannier basis representation of our system is obtained through a unitary transformation of the Bloch basis wavefunctions \cite{wannierOG,kohnWannier,brown2023realspace}. The Wannier basis describes a periodic system using a basis of states known as Wannier functions localized about individual lattice sites. The unitary transformation used to obtain the Wannier basis is not unique and we use the formalism developed in \cite{kohnWannier} to obtain a basis of maximally localized Wannier functions \cite{RevModPhys.84.1419}. With this, the Wannier function in band $m$ localized about lattice site $R$ for our system is related to the Bloch basis wavefunctions through the following relation:

\begin{align}
	\ket{w_{m, R}} &= \frac{1}{\sqrt{N}} \sum_{k} e^{- i k R} \ket{\psi_{m, k}}, \label{eq:bloch2wannier} \\
	\ket{\psi_{m, k}} &= \frac{1}{\sqrt{N}} \sum_{R} e^{i k R} \ket{w_{m, R}} . \label{eq:wannier2bloch}
\end{align} 

Operators in the Wannier basis couple Wannier functions in each band separated by $\Delta R = n a_0$ (integer $n$) lattice sites such that any operator $\hat{\mathcal{O}}$ may be expressed as a summation of operators coupling Wannier functions separated by all possible separations $\Delta R$:

\begin{align}
	\hat{\mathcal{O}} &= \sum_{\Delta R} \hat{\mathcal{O}}^{\Delta R} \label{eq:operatorWannierTotal}, \\
	\hat{\mathcal{O}}^{\Delta R} &= \sum_{m, m'} \sum_{R} \mathcal{O}_{m, m'}^{\Delta R} \ket{w_{m, R}} \bra{w_{m', R - \Delta R}}, \label{eq:operatorWannierDR} \\
	\mathcal{O}_{m, m'}^{\Delta R} &= \frac{1}{N} \sum_{k} e^{i k \Delta R} \mathcal{O}_{m, m'}^{k} \label{eq:operatorWannierME}
\end{align}

\noindent
where $\mathcal{O}_{m, m'}^{k}$ is the matrix element of $\hat{\mathcal{O}}$ in the Bloch basis between states in bands $m$ and $m'$ with crystal momentum $k$. 

We propagate the system in time using the density matrix formalism with the initial condition of a full (empty) valence (conduction) band and describe the interaction with the external field using the length gauge such that the equation of motion for the density matrix element $\rho_{m, m'}^{R, R'}(t)$ is given as follows:

\begin{equation}
	\begin{split}
		i \frac{\partial \rho_{m, m'}^{R, R'}}{\partial t} = \left[ \hat{H}_0 + F(t) \hat{d} , \hat{\rho}(t) \right] \vphantom{x}_{m, m'}^{R, R'}  + i w_{m, m'}^{R, R'} \rho_{m, m'}^{R, R'} (t),
		\label{eq:eom}
	\end{split}
\end{equation}

\noindent 
where $\hat{H}_0$ and $\hat{d}$ are the field-free Hamiltonian and dipole operators in the Wannier basis calculated from their Bloch basis representations using Eqs. (\ref{eq:operatorWannierTotal}-\ref{eq:operatorWannierME}) and we have incorporated dephasing through $w_{m, m'}^{R, R'}$. 

The dephasing term $w_{m, m'}^{R, R'}$ consists of two components. First, we include a dephasing time $T_2$ which describes microscopic dynamics beyond the scope of the single-active electron approximation and leads to a suppression of interband coherence. We also include the spatial-dephasing mechanism first proposed in \cite{brown2023realspace}, which preferentially suppresses interband coherence between distantly separated Wannier functions through a term $\mathcal{W}_{m, m'}^{R, R'}$ such that $w_{m, m'}^{R, R'}$ is defined as follows:

\begin{equation}
	w_{m, m'}^{R, R'} = - \left( 1 - \delta_{m, m'} \right) \left( \frac{1}{T_2} + \mathcal{W}_{m, m'}^{R, R'} \right) .
\end{equation}

\noindent
We define $\mathcal{W}_{m, m'}^{R, R'}$ such that the spatial-dephasing rate increases quadratically with separation $\Delta R = | R - R' |$ above a threshold separation $\Delta R_0$ with strength $\gamma$:

\begin{equation}
	\mathcal{W}_{m, m'}^{R, R'} = \begin{cases} \gamma \left( R - R' \right)^2 & | R - R' | \ge \Delta R_0 \\ 0 & | R - R' | < \Delta R_0 \end{cases} . 
	\label{eq:spatialDephasing}
\end{equation}

\begin{figure*}
	\centering
	\includegraphics[width=\textwidth]{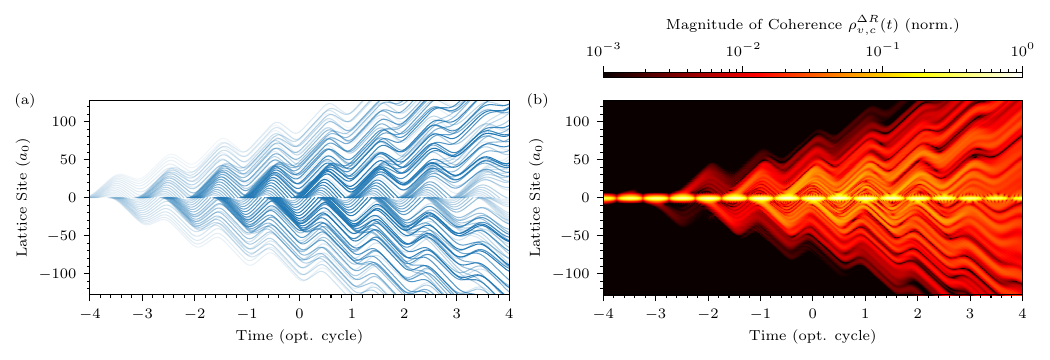}
	\caption{(a) Electron-hole wave packet trajectories calculated from the semi-classical recollision model \cite{vampa1} and (b) the coherence between the valence band Wannier function at lattice site $0$ and the conduction band Wannier functions at lattice sites $\Delta R$ during the interaction with the driving laser field calculated using the model described in Section \ref{section:NumericalModel}. The driving field is an eight-cycle Gaussian driving pulse with peak intensity $3.1 \times 10^{11}$ W/cm$^{2}$ and wavelength 3.2 \textmu m. The transparency of a given trajectory in (a) is determined by the normalized square of the driving electric field intensity and the numerical simulation in (b) is performed with no dephasing.}
	\label{fig:semiclassicalAndNoDephasingTraj}
\end{figure*}

After solving for $\hat{\rho}(t)$ using Eq. (\ref{eq:eom}-\ref{eq:spatialDephasing}), we calculate the current between Wannier functions in bands $m$ and $m'$ separated by $\Delta R$ lattice sites as follows:

\begin{equation}
	j_{m, m'}^{\Delta R} (t) \propto p_{m, m'}^{- \Delta R} \rho_{m', m}^{\Delta R} (t),
	\label{eq:currentDR}
\end{equation}

\noindent
where $p_{m, m'}^{- \Delta R}$ is obtained from the Bloch basis momentum operator matrix elements using Eq. (\ref{eq:operatorWannierME}), where we have set $\rho_{m', m}^{\Delta R}(t) = \rho_{m', m}^{0, \Delta R}(t)$ without a loss of generality since $\mathcal{O}_{m', m}^{R, R + \Delta R} = \mathcal{O}_{m', m}^{0, \Delta R}$ due to the periodicity of the system. The emitted spectrum is then calculated from $j_{m, m'}^{\Delta R}(t)$ as

\begin{equation}
	\tilde{s}_{m, m'}^{\Delta R} \left(\Omega\right) = - i \Omega \int_{- \infty}^{\infty} f(t) j_{m, m'}^{\Delta R}(t) e^{i \Omega t} dt,
	\label{eq:spectrumDR}
\end{equation}

\noindent
where $f(t)$ is a Gaussian window function. The total current is obtained by summing Eq. (\ref{eq:currentDR}) over all band indices and separations.

As we will show in Sections \ref{section:recollisionAndWannier} and \ref{section:results}, the density matrix elements $\rho_{m, m'}^{\Delta R} (t)$ can be used to visualize recollision dynamics in real-space. For microscopic simulations, this can be done through the calculated density matrix elements directly from the solution to Eq. (\ref{eq:eom}). For macroscopic simulations, however, we calculate the \emph{observed} density matrix elements through the observed current in the far-field. To accomplish this, we first calculate the separation- and band-resolved HHG emission $\tilde{s}_{m, m'}^{\Delta R} (r, \Omega)$ at radius $r$ across a radially symmetric Gaussian beam using Eqs. (\ref{eq:bloch2wannier}-\ref{eq:spectrumDR}). We then propagate the resultant HHG beam fronts into the far-field, from which we calculate the far-field time-dependent current. We then calculate the \emph{observed} matrix elements which survive far-field propagation from the on-axis far-field time-dependent current from Eq. (\ref{eq:currentDR}). Details of our macroscopic propagation model can be found in Appendix \ref{section:macroscopicPropagation}.

\section{Recollision Dynamics and the Wannier Basis}
\label{section:recollisionAndWannier}

Recent works \cite{brown2023realspace,PhysRevB.100.195201} have argued the advantages of using the Wannier basis for simulations of strong-field driven dynamics in solid-state systems. Here, we expand upon this and show the relationship between the Wannier basis and recollision trajectories can be made explicit if we consider the semi-classical model of recollision. We consider a two-band system (valence $v$ and conduction $c$) and assume negligible depletion of the ground state. With this and a continuous crystal momentum, an analytic expression for the density matrix element $\rho_{c,v}^{\kappa(t)}(t)$ in the Houston basis \cite{hhgFromBEiS} with $\kappa(t) = k_0 + A(t)$ where $k_0$ is the initial crystal momentum of the state is given as follows: 

\begin{equation}
	\begin{split}
		\hat{\rho}_{c,v}^{\kappa(t)}(t) &= i \int_{- \infty}^{t} e^{- i S[\kappa] (t_b, t_r)} F(t_b) d_{c,v}[\kappa] (t_b) dt_b . 
	\end{split}
	\label{eq:houstonMatrixElement}
\end{equation}

\noindent
The corresponding matrix element in the length gauge is obtained by simply mapping $\kappa(t) \to k$ \cite{Yue:22}. We then perform the summation over $k$ in Eq. (\ref{eq:operatorWannierME}) as an integral to transform Eq. (\ref{eq:houstonMatrixElement}) to the Wannier basis (now including $k$ in the function arguments):

\begin{equation}
	\rho_{c, v}^{\Delta R} (t) \propto \int dk \int dt_b e^{- i S(k, t_b, t_r) + i k \Delta R} F(t_b) d_{c,v}(k, t_b).
\end{equation}

\noindent
Applying the saddle-point approximation to the integration over $k$ yields the following saddle-point equation:

\begin{equation}
	0 = \frac{\partial S}{\partial k} - \Delta R ,
\end{equation}

\noindent
where $\partial S / \partial k$ is the displacement condition from the semi-classical recollision model \cite{vampa1}. That is, the dominant contribution to the density matrix element $\rho_{c, v}^{\Delta R}(t)$ occurs when the semi-classical electron-hole displacement is equal to $\Delta R$. 

Fig. \ref{fig:semiclassicalAndNoDephasingTraj} shows (a) the recollision trajectories calculated from the semiclassical model \cite{vampa1} and (b) the amplitude of the coherence $\rho_{v, c}^{\Delta R}(t)$ between the valence band $v$ Wannier function at lattice site $R = 0$ and the conduction band $c$ Wannier functions at lattice sites $\Delta R$ during the interaction with the driving laser field calculated from the model described in Section \ref{section:NumericalModel}. The driving field is an eight-cycle Gaussian pulse with peak intensity of $3.1 \times 10^{11}$ W/cm$^{2}$ and central wavelength 3.2 \textmu m. The transparency of the recollision trajectories in (a) are determined by the normalized square of the driving electric field and the numerical simulation shown in (b) is calculated without dephasing. The dynamics depicted in (b) are nearly identical to those depicted in (a), demonstrating the utility of the Wannier basis for \emph{ab initio} investigations into real-space recollision dynamics. 

\section{Recollision Trajectories, Dephasing, and Far-Field Propagation}
\label{section:results}

\begin{figure*}
	\centering
	\includegraphics[width=0.95\textwidth]{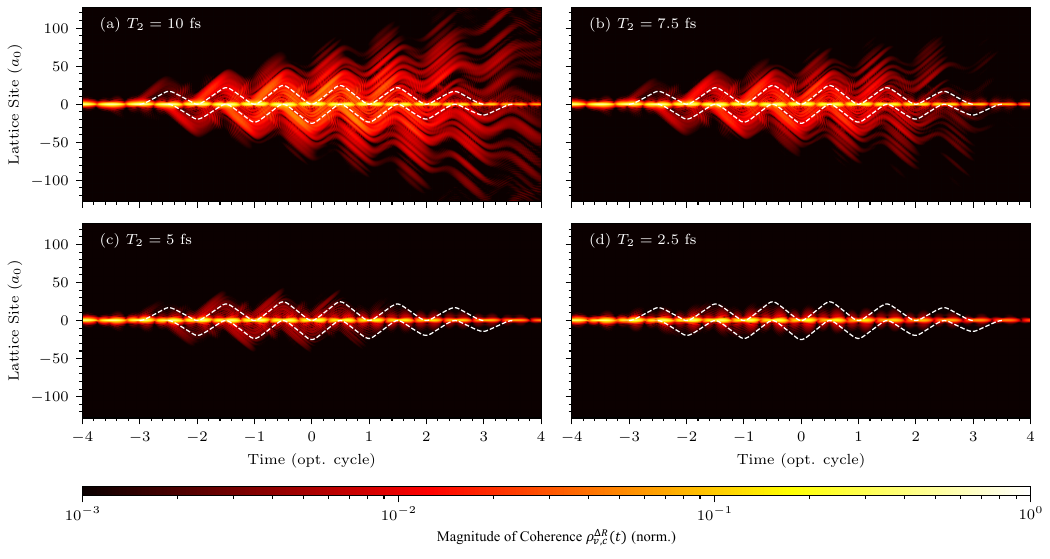}
	\caption{Magnitude of the density matrix element $\rho_{v, c}^{0, R}(t)$ depicting the coherence between the valence band Wannier orbital at lattice site 0 and the conduction band Wannier orbitals separated at lattice sites $R$ as a function of time calculated using dephasing times of (a) 10 fs, (b) 7.5 fs, (c) 5 fs, and (d) 2.5 fs. The overlaid dashed white lines in each subplot denote the trajectory with the furthest excursion distance which recombines with its parent lattice site. The driving field is an eight-cycle Gaussian pulse with wavelength 3.2 \textmu m and a peak intensity of $3.1 \times 10^{11}$ W/cm$^{2}$.}
	\label{fig:T2Traj}
\end{figure*}

We are now ready to investigate the relationship between real-space recollision dynamics, dephasing, and macroscopic propagation. We begin by presenting the results from our trajectory analysis from microscopic simulations of HHG calculated with various uniform dephasing times $T_2$ and no spatial-dephasing. Fig. \ref{fig:T2Traj} shows the magnitude of the density matrix elements $\rho_{v, c}^{\Delta_R}(t)$ during the interaction with the driving field calculated with dephasing times of (a) 10 fs, (b) 7.5 fs, (c) 5 fs, and (d) 2.5 fs. The driving field is an eight-cycle Gaussian pulse with wavelength 3.2 \textmu m and a peak intensity of $3.1 \times 10^{11}$ W/cm$^{2}$. In each subplot, the recollision trajectory with the maximum excursion distance which enters the conduction band at and returns to its parent lattice site (i.e. which tunnels at the peak of the field) calculated using the semiclassical model is shown by the overlaid white lines. 

The results calculated with a dephasing time of 10 fs depicted in Fig. \ref{fig:T2Traj} (a) are virtually identical to the semiclassical trajectories depicted in Fig. \ref{fig:semiclassicalAndNoDephasingTraj} (a) and the numerical result in Fig. \ref{fig:semiclassicalAndNoDephasingTraj} (b) calculated without dephasing. Due to the long dephasing time used, coherences between lattice sites with separations $\sim 100$ lattice sites are populated by trajectories which tunnels before the peak of a given half-cycle of the driving field. In gas-phase HHG, HHG emission is localized at the position of the parent ion and occurs when the free electron recombines into the ground state \cite{PhysRevA.49.2117}. Thus, the components of the free electron wave packet which never return to their parent ion negligibly contribute to calculated HHG spectra. In contrast, the cross-section for interband HHG emission is appreciable for electron-hole separations across several lattice sites \cite{WQC} and intraband HHG emission occurs throughout the trajectory of the electron-hole wave packet prior to recombination. Consequently, the components of the electron-hole wave packet which never return to their parent lattice site are much more significant than their analogue in gas-phase HHG. 

As the dephasing time is decreased, the electron-hole wave packet does not have sufficient time to populate these distant coherences before being suppressed. This is clearly evident in Fig. \ref{fig:T2Traj} (b, c, d), where the maximum excursion distance of the electron-hole wave packet is is reduced as the dephasing time is decreased. The coherence in the vicinity of the parent lattice site is, however, similarly suppressed. Since the timescale of recollision trajectories is typically on the order of one-quarter optical cycle, dephasing times $\sim 1$ fs lead to the suppression of trajectories which do recombine and contribute to experimentally measured HHG spectra. 

\begin{figure*}
	\includegraphics[width=\textwidth]{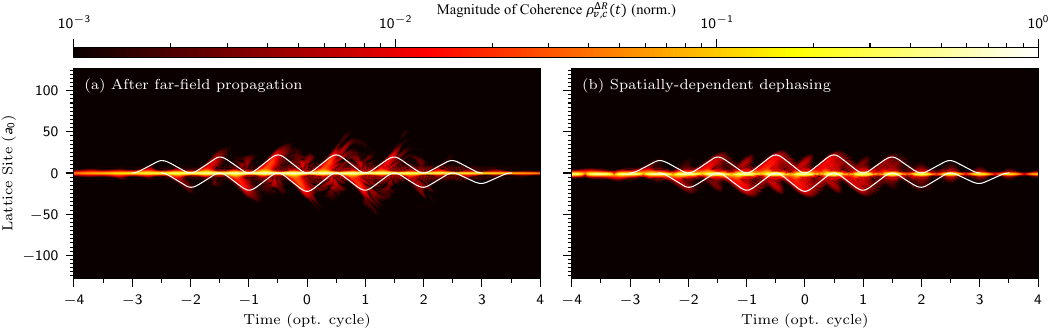}	
	\caption{The magnitude of the coherence elements between lattice sites of the density matrix calculated from the time-dependent current obtained from (a) the inverse Fourier transform of the on-axis lattice site separation resolved near-field HHG spectrum after spatial filtering in the far-field and (b) a microscopic simulation using the spatially-dependent dephasing mechanism described in \cite{brown2023realspace}. The time-dependent field is the same as used to generate the dynamics depicted in Fig. \ref{fig:T2Traj}. The macroscopic simulation is performed using by calculating HHG emission at the focus of a a radially symmetric Gaussian beam with beam waist 50 \textmu m and propagating a distance of $1$ m into the far-field.}
	\label{fig:trajFF}
\end{figure*}

Since the use of extremely short dephasing times stems from the requirement to obtain agreement between simulated and experimentally measured HHG spectra, we now calculate the dynamics of the electron-hole wave packet which contribute to HHG spectra in experimental conditions by performing a macroscopic simulation. To do this, we consider a driving electric field with the same temporal profile as that used to obtain the results presented in Figs. \ref{fig:semiclassicalAndNoDephasingTraj} and \ref{fig:T2Traj} and calculate HHG emission from a monolayer of our system placed at the focus of the driving electric field beam front, which is modelled as a Gaussian beam with a waist of 50 \textmu m. We then propagate the resultant HHG emission 1 m into the far-field and use the on-axis HHG spectrum to obtain the \emph{observed} coherence between Wannier functions, as described in Section \ref{section:NumericalModel}.

Fig. \ref{fig:trajFF} (a) shows the lattice site separation-resolved valence-coherence band coherence displayed as in Fig. \ref{fig:T2Traj} extracted from the far-field HHG spectrum. As in Fig. \ref{fig:T2Traj}, we overlay the recollision trajectories with the maximum excursion distance which return to their parent lattice site by a white line and the results are shown on the same colour-scale as those in \ref{fig:T2Traj} for clarity. From Fig. \ref{fig:trajFF} (a) we immediately observe that far-field propagation significantly suppresses contributions to HHG spectra arising from distantly separated coherences but retains those whose separations are comparable to that predicted by the semi-classical recollision model. As we will argue below, this is due to the innate intensity-dependent phase characteristic of non-perturbative HHG emission.

One of the features characteristic of non-perturbative harmonic generation is the strong intensity-dependence of the spectral phase in non-perturbative HHG \cite{doi:10.1098/rsta.2017.0468}. This intensity-dependent phase gives rise to an innate curvature of the spectral beam fronts in HHG \cite{Frumker:12,Abadie:18}. In atomic HHG, this intensity-dependent phase is particularly important when using long wavelength driving lasers, where the large intensity-dependence of the phase of HHG emission from long trajectories reduces their importance in experimentally measured spectra \cite{Zhang2016}. Accordingly, these macroscopic effects arising from experimental conditions are incorporated into microscopic simulations of atomic HHG through the use of complex absorbing potentials which suppress HHG emission from trajectories whose real-space excursion results in a large intensity-dependence of their phase in order to obtain agreement between simulated and experimentally measured spectra \cite{Rothhardt_2014}.

We now make a direct link between microscopic recollision dynamics in solid-state HHG and the macroscopic propagation of HHG emission. One distinguishing aspect of solid-state recollision in periodic systems when compared with atomic systems is the possibility for transitions between Wannier functions separated by large distances \cite{WQC}. In the presence of a strong field, however, there will be an induced voltage throughout the lattice which introduces an energy differential between lattice sites \cite{PhysRevB.90.085313}. If we consider a periodic system in the presence of an external field vector potential $A(t)$, we may estimate the effect of this field-induced voltage on the phase $\Phi(\Delta R)$ of an electron trajectory which tunnels at time $t_b$ and recombines at time $t_r$ across a distance of $\Delta R$ as follows:

\begin{equation}
	\Phi\left(\Delta R\right) = - \left( A(t_r) - A(t_b) \right) \Delta R. 
	\label{eq:phaseDR}
\end{equation}

For realistic system and experimental parameters, this phase exceeds $2 \pi$ even for relatively small separations (e.g. $\Phi(4 a_0) > 2 \pi$ for the parameters considered here). Under experimental conditions, the phase in Eq. (\ref{eq:phaseDR}) will exhibit a large wavefront curvature due to the varying intensity profile of the driving field beam front and the associated HHG emission will be highly divergent. Consequently, the influence of HHG emission from radiative transitions across large distances on experimentally measured HHG spectra will be diminished when compared with their influence on microscopically calculated spectra similarly to the diminished importance of HHG emission from long trajectories in atomic HHG when long wavelength driving lasers are used.  

In light of this, we now adopt an approach analogous to that used in atomic HHG simulations with long wavelength driving lasers for the suppression of HHG emission from long trajectories. We perform a microscopic simulation using a spatial-dephasing mechanism as first proposed in \cite{brown2023realspace} to preferentially suppress HHG emission from transitions between distantly separated Wannier functions. We use a dephasing time $T_2 = 10$ fs and the spatial-dephasing boundary given by Eq. (\ref{eq:spatialDephasing}) with  $\Delta R_0 = 10 a_0$ and strength parameter $\gamma$ such that an effective dephasing time of 2 fs is obtained for separations $\Delta R = 24 a_0$. The coherence $\rho_{v, c}^{\Delta R}(t)$ obtained from this simulation is shown in Fig. \ref{fig:trajFF} (b). Like the result obtained after simulating far-field propagation in Fig. \ref{fig:trajFF} (a), the spatial-dephasing mechanism preferentially suppresses distantly separated coherences and retains similar components of the electron-hole wave packet to the far-field result. This is unsurprising, given that it has been shown that the use of a spatial-dephasing mechanism yields HHG spectra which are nearly indistinguishable from those obtained after far-field propagation without the use of ultrafast dephasing times \cite{brown2023realspace}.

The use of a spatial-dephasing mechanism is analogous to the use of an absorbing boundary in simulations of atomic HHG to suppress components of the electron-hole wave packet corresponding to trajectories whose emission phase exhibits a large intensity-dependence, albeit with differing physical justifications. While the components of the continuum electron wave packet which are suppressed in simulations of atomic HHG with long wavelength driving lasers return to their parent lattice ion, their innate divergence due to their intensity-dependent phase reduces their relative importance in experimentally measured HHG spectra. In solid-state HHG, there is an additional class of trajectories which never return to their parent lattice site but still lead to HHG emission. Such trajectories significantly contribute to microscopically simulated HHG spectra, but the significance of their emission in experimentally measured HHG spectra is reduced due to the intensity-dependent phase arising from the field-induced voltage across a lattice. In both cases, however, the artificial suppression of the electron-hole wave packet is introduced to mimic macroscopic effects in microscopic simulation. While the use of absorbing boundaries to incorporate macroscopic effects in microscopic simulation is ubiquitous throughout the field of gas-phase HHG, such an approach has, until recently, not been attempted in solids. 

Since spatial-dephasing is motivated by the intensity-dependence of the phase of HHG emission, it is generally applicable to simulations of solid-state HHG in all materials undergoing recollision which exhibit significant interband HHG emission. Our the identification of spatial-dephasing, thus, resolves the longstanding issue regarding the seeming universality of the requirement for ultrafast dephasing times. The generality of the intensity-dependent phase in HHG means that spatial dephasing must be accounted for in all studies of dephasing and carrier dynamics in condensed matter performed using HHG spectroscopy. This will be particularly important for systems which are expected to exhibit microscopic dynamics leading to dephasing on the femtosecond timescale, such as disorder \cite{Orlando:20} or many-body interactions \cite{PhysRevLett.112.257402}. 

\section{Conclusion}

By using a real-space model of solid-state HHG, we have established critical links between atomic and solid-state HHG. Like atomic HHG, where HHG spectra are dominated by recollision trajectories which return to their parent ion, far-field solid-state HHG spectra are predominantly shaped by trajectories which return to the vicinity of their parent lattice site. The widespread use of ultrafast dephasing times $\sim 1$ fs, however, artificially suppresses recollision trajectories which return to their parent lattice site and contribute to far-field HHG spectra. 

When a spatial dephasing mechanism designed to preferentially suppress HHG emission from transitions between distantly separated Wannier functions is used, the dynamics obtained from microscopic simulations quantitatively agree with results obtained after simulations of far-field propagation calculated without ultrafast dephasing times. The suppression of HHG emission from transitions between distantly separated states is motivated by the large intensity-dependence of their phase of emission and is completely analogous to the use of complex absorbing potentials to suppress HHG emission from long trajectories in simulations of atomic HHG using long driving laser wavelengths in that both methods serve to incorporate macroscopic effects which shape experimentally measured HHG spectra into microscopic simulation. Our results will be important for the interpretation of all solid-state HHG experiments studying ultrafast carrier dynamics in condensed matter using solid-state HHG spectroscopy. 

G. G. Brown acknowledges funding from the European Union’s Horizon 2020 research and innovation programme under grant agreement No. 899794 (Optologic). M. Ivanov  acknowledges funding from the SFB 1477 ``Light Matter Interaction at Interfaces" project number 441234705. \'{A}.J.G. acknowledges funding from the European Union’s Horizon 2020 research and innovation programme under the Marie Skłodowska-Curie grant agreement no. 101028938. \'{A}.J.G. acknowledges funding from the European Union’s Horizon 2020 research and innovation programme under the Marie Skłodowska-Curie grant agreement no. 101028938. R. E. F. Silva acknowledges support from the fellowship LCF/BQ/PR21/11840008 from ``La Caixa” Foundation (ID 100010434). We thank A. Marini, H. Gross, E. Goulielmakis, and A. Leitenstorfer for exceptionally useful comments.

\appendix

\section{Macroscopic Propagation}
\label{section:macroscopicPropagation}

For our macroscopic simulations, we calculate HHG emission using Eqs. (\ref{eq:bloch2wannier}-\ref{eq:spectrumDR}) across a radially symmetric Gaussian beam with time $t$-dependent electric field at radial position $r$, $F(r, t)$: 

\begin{equation}
	F(r, t) = F_0 e^{- (r / \sigma)^2} e^{- (2 t / \tau_0)^2} \cos\left(\omega_0 t + \varphi\right), 
\end{equation}

\noindent 
where $F_0$ is the peak electric field amplitude, $\sigma$ is the Gaussian beam waist at the focus,  $\omega_0$ is the carrier frequency of the driving field, $\tau_0$ is the pulse duration, and $\varphi$ is the carrier-envelope phase (CEP). 

We calculate HHG emission from a two-dimensional surface placed in the focal plane of the driving field. Using Eq. (\ref{eq:currentDR}) and (\ref{eq:spectrumDR}), we obtain the HHG spectra emitted from transitions between Wannier functions in bands $m$ and $m'$ separated by $\Delta_R$ lattice sites at each considered radial position of the driving field, $\tilde{s}^{\Delta_R}_{m,m'}(r, \Omega)$. We then propagate the near-field spectra into the far-field by first evaluating the radially symmetric Fourier transform (i.e. the Hankel transform) of the near-field HHG emission, which we denote as $\tilde{\mc{J}}_{m, m'}^{\Delta_R} (k_r, \Omega)$:

\begin{equation}
	\begin{split}
		\tilde{\mc{J}}^{\Delta_R}_{m, m'}\left( \kappa_r , \Omega \right) &= 2 \pi \int_0^{\infty} \tilde{s}^{\Delta_R}_{m,m'}\left(r, \Omega\right) J_0 \left( r \kappa_r \right) r dr ,
	\end{split}
	\label{eq:hankelFWD}
\end{equation}

\noindent 
where $J_0(r \kappa_r)$ denotes the zeroth-order Bessel function of the first kind and $\kappa_r$ denotes the radial spatial frequency related to the Hankel transform. We let $U_{\Omega}(z)$ represent the far-field propagator for frequency $\Omega$ for a propagation distance $z$:

\begin{equation}
	U_{\Omega}(z, \kappa_r) = e^{i \overline{\kappa}_{\Omega} z}, \label{eq:zPropagator} , 
\end{equation}

\noindent 
where $\overline{\kappa}_{\Omega} = \sqrt{\kappa_{\Omega}^2 - k_r^2}$, $\kappa_{\Omega} = \Omega / c$, and $c$ is the speed of light. The HHG beam at position $z$ is then calculated by applying the propagator in Eq. (\ref{eq:zPropagator}) and calculating the inverse Hankel transform:

\begin{equation}
	\begin{split}
		\tilde{s}^{\Delta_R}_{m,m'}\left(r, z, \Omega\right) &= 2 \pi \int_0^{\infty} U_{\Omega}(z,\kappa_r) \tilde{\mc{J}}^{\Delta_R}_{m, m'}\left( \kappa_r , \Omega \right) \\
		&\hspace{5 mm} \times J_0 \left( \kappa_r r \right) \kappa_r d\kappa_r,
	\end{split}
	\label{eq:spectrumZ}
\end{equation}

The current $j_{m, m'}^{\Delta R}(r, z, t)$ at radius $r$ after propagating a distance $z$ into the far-field at time $t$ is then obtained Eq. (\ref{eq:spectrumZ}) using an inverse Fourier transform. Fig. \ref{fig:spectraFF} depicts the radially integrated (blue) near- and (red) far-field HHG spectra calculated in the focal plane of a Gaussian beam with a waist of 100 \textmu m, propagated a distance of 1 m into the far-field, and with a dephasing time of 10 fs. Similarly to the results in \cite{Abadie:18}, the near-field HHG spectrum exhibits a chaotic structure, whereas the far-field spectrum exhibits a well-defined harmonic structure. 

The \emph{observed} density matrix elements $\rho_{m, m'}^{\Delta R}(r, z, t)$ are then calculated from the far-field current $j_{m, m'}^{\Delta R}(r, z, t)$ using Eq. (\ref{eq:currentDR}):

\begin{equation}
	\rho_{m, m'}^{\Delta R} (r, z, t) = \frac{j_{m, m'}^{\Delta R}(t)}{p_{m', m}^{\Delta R}} . 
\end{equation}

\bibliography{apssamp}

\end{document}